\documentclass[preprint,aps,tightenlines,floatfix]{revtex4}
\usepackage{epsfig}
\usepackage{graphics}
\usepackage{amsmath}

\begin{document}

\preprint{
\vbox{
\hbox{JLAB-THY-03-04}
\hbox{OUTP-03-03P}
}}

\title{Symmetry Breaking and Quark-Hadron Duality \\
	in Structure Functions}

\author{F.E. Close}
\affiliation{Department of Theoretical Physics,
	University of Oxford,
	Keble Rd.,
	Oxford, OX1 3NP, England}
\author{W. Melnitchouk}
\affiliation{Jefferson Lab,
	12000 Jefferson Avenue,
	Newport News, VA 23606\\ \\}

\begin{abstract}
We identify conditions under which a summation over nucleon resonances
can yield, via quark-hadron duality, parton model results for
electromagnetic and neutrino structure functions at large $x$.
While a summation over the lowest even and odd parity multiplets is
sufficient to achieve duality in the symmetric quark model, a
suppression of transitions to specific final states is required for
more realistic cases incorporating SU(6) breaking.
We outline several scenarios consistent with duality, discuss their
implications for the high $Q^2$ behavior of transition form factors,
and illustrate how they can expose the patterns in the flavor-spin
dependence of inter-quark forces.
\end{abstract}

\maketitle

\section{Introduction}

The relation between resonances and deep inelastic structure functions
has been the subject of considerable interest recently.
This has been partly prompted by recent high precision data from
Jefferson Lab \cite{F2P} on the unpolarized $F_2$ structure function of
the proton in the resonance region, which showed a striking similarity,
when averaged over resonances, to the structure function measured at
much higher energies in the deep inelastic region.
This phenomenon was first observed some time ago by Bloom and Gilman
\cite{BG}, who found that when integrated over the mass of the inclusive
hadronic final state, $W$, the scaling structure function at high $Q^2$,
$F_2^{\rm scaling}$, smoothly averages that measured in the region
dominated by low-lying resonances, $F_2^{\rm exp}$,
\begin{eqnarray}
\int dW\ F_2^{\rm exp}(W^2,Q^2) &=& \int dW\ F_2^{\rm scaling}(W^2/Q^2)\ .
\label{bgdual}
\end{eqnarray}
The integrand on the left-hand-side of Eq.~(\ref{bgdual}) represents the
structure function in the resonance region at low $Q^2$, while that on
the right-hand-side corresponds to the scaling function, measured in the
deep inelastic region at high $Q^2$.
The latter is described by leading twist, perturbative QCD, as an
incoherent sum over quark flavors, $\sum e_i^2$; the former involves
coherent excitation of resonances.

Global duality is said to hold after integration over all $W$ in
Eq.~(\ref{bgdual}).
This equality can be related to the suppression of higher twist
contributions to moments of the structure function \cite{RUJ}, in which
the total moment becomes dominated by the leading twist ($\approx Q^2$
independent) component at some lower value of $Q^2$.
Information on all coherent interaction dynamics is subsequently lost.
A more local form of duality is also observed \cite{F2P}, in which the
equality in Eq.~(\ref{bgdual}) holds for restricted regions of $W$
integration --- specifically, for the three prominent resonance regions
at $W \alt 1.8$~GeV.
The duality between the resonance and scaling structure functions is also
being investigated in other structure functions, such as the longitudinal
structure function \cite{F1P}, and spin dependent structure functions of
the proton and neutron \cite{G1P,G1N}.
For spin-dependent structure functions, in particular, the workings of
duality are more intricate, as the difference of cross sections no longer
needs to be positive.
An example is the contribution of the $\Delta$ resonance to the $g_1$
structure function of the proton, which is large and negative at low
$Q^2$, but may become positive at higher $Q^2$.

Early work within the SU(6) symmetric quark model \cite{CGK,CGNU,COT}
showed how the ratios of various deep inelastic structure functions at
$x = Q^2/2M\nu \sim 1/3$, both spin dependent and independent, could be
dual to a sum over $N^*$ resonances in the SU(6)$^P$={\bf 56}$^+$ and
{\bf 70}$^-$ representations.
With the emergence of precision data, showing detailed and interesting
$x$ dependence as $x \to 1$, various questions arise:
\begin{itemize}

\item How do changes in ratios as $x \to 1$ relate to the pattern of
$N^*$ resonances identified in the quark model \cite{CGK,CGNU,COT}?

\item Are certain families (spin-flavor correlations) of resonances
required to die out at large $Q^2$ in order to maintain duality?
If so, can electroproduction of specific examples of such resonances
test this?

\item Can such a program reveal the flavor-spin dependence of short
distance forces in the QCD bound state?

\end{itemize}

The aim of this paper is to make a first orientation towards answering
such questions.
Quark models based on SU(6) spin-flavor symmetry provide benchmark
descriptions of baryon spectra, as well as transitions to excited $N^*$
states.
To allow the origins of duality to remain manifest throughout our
discussion, we shall restrict ourselves to the framework of the
quark model, but consider the effects of SU(6) breaking explicitly.
Such models serve as convenient laboratories for examining the
generality of the quark-hadron duality phenomenon in more realistic
scenarios than in earlier discussions.
The duality between the simplest SU(6) quark parton model results
\cite{KW} for ratios of structure functions with sums over the
{\bf 56}$^+$ and {\bf 70}$^-$ coherent $N^*$ excitations was described
in Refs.~\cite{CGK,CGNU,CG}.
An essential feature of those analyses was that SU(6) was exact and that
exotics in the $t$-channel were suppressed.
In a global sense such results are self-consistent: the absence of
$t$-channel exotics equates with an absence of
$\gamma \gamma \to qq \bar{q}\bar{q}$ couplings, and hence, in effect,
to the presence only of incoherent diagrams where the photons couple to
the same quark, $\gamma q \to \gamma q$ (see Fig.~1(a)).

Although the $s$-channel sum was shown to be dual for ratios of
incoherent quantities \cite{CGK,CGNU,COT}, this alone did not explain
why (or if) any individual sum over states scaled.
%
%
Recently, the transition from resonances to scaling has been explored
in microscopic models at the quark level.
The phenomenological quark model duality of Refs.~\cite{CGK,CGNU,COT}
was recently shown \cite{CI} to arise in a simple model of spinless
constituents.
A model in which the hadron consisted of a point-like scalar ``quark''
bound to an infinitely massive core by a harmonic oscillator potential
was used \cite{IJMV} to explicitly demonstrate how a sum over infinitely
narrow resonances can lead to a structure function which scales in the
$Q^2 \to \infty$ limit. 
These ideas have been further developed \cite{OTHER} to give an
increasingly solid model underpinning of this phenomenological duality.

Since the original quark model predictions were made in the 1970's, the
quantity and quality of structure function data have improved
dramatically.
We now know, for instance, that in some regions of $x$ SU(6) symmetry is
badly broken, with the strongest deviations from the naive SU(6)
expectations being prevalent at large values of $x$.
The new data will set challenges for theories of quark-hadron duality.
There are critical questions which now need to be addressed, such as:

\begin{itemize}

\item
Can duality survive locally in $x$, and what do the observed variations
in $x$ require of $N^*$ excitations if duality is to survive?

\item
What families (spin flavor correlations, or SU(6) multiplets) are
suppressed as $x \to 1$, or equivalently $Q^2 \to \infty$, for duality
to hold?

\item
Does the excitation of low lying prominent $N^*$ resonances, belonging
to such families, exhibit such behavior?

\end{itemize}

If the $x \to 1$ systematics for $N^*$ families are not matched by
specific $N \to N^*$ transition form factors as a function of $Q^2$,
then duality fails.
If, however, they do match, then this can expose the patterns in the
flavor-spin dependence of inter-quark forces.

In this paper we explore the question of whether quark-hadron duality
exists in structure functions for the more realistic scenario in which
SU(6) is explicitly broken.
We focus on both electromagnetic and neutrino scattering.
While most of the phenomenological information comes from electron
scattering, neutrino-induced reactions provide an important consistency
check on the derived duality relations, and predictions for neutrino
structure function ratios can be tested once high-intensity neutrino
beams become available \cite{NUMI}.
After reviewing the symmetric SU(6) quark model results for structure
functions in Section~II, we identify in Section~III the necessary
patterns of $N \to N^*$ suppression in order to obtain structure
functions which are compatible with data and expectations from hard
scattering at large $x$ and higher $Q^2$.
In the process we derive duality relations for various structure function
ratios, in which the breaking of SU(6) symmetry is parameterized in terms
of $x$-dependent mixing angles.
Fixing the mixing angles by the unpolarized neutron to proton structure
function ratio data then allows us to make explicit predictions for the
$x$-dependence of polarization asymmetries for the proton and neutron,
under various symmetry breaking scenarios.
Experimental signatures for the corresponding $N \to N^*$ suppressions are
discussed in Section~IV, and conclusions and ideas for future developments
summarized in Section~V.

\section{Duality and the Quark Model}

The SU(6) spin-flavor symmetric quark model serves as a useful basis in
which one may visualize both the principles underpinning the phenomenon
of quark-hadron duality and at the same time provide a reasonably close
contact with phenomenology.
Following earlier work in Refs.~\cite{CGK,CGNU,COT,CG,CKO}, it was shown
by Close \& Isgur \cite{CI} that the structure function ratios of the
symmetric quark model can be obtained by summing over appropriate sets
of baryon resonances.
Higher twist effects, which give rise to violations of duality through
non-diagonal quark transitions, such as in the ``cat's ears'' diagram in
Fig.~1(b), can be shown to cancel in a small energy range appropriate for
summing over neighboring odd and even parity states.
In the SU(6) quark model this corresponds to summing over states in the
{\bf 56}$^+$ ($L=0$) and {\bf 70}$^-$ ($L=1$) multiplets, with each
representation weighted equally.
The spin-averaged transverse $F_1$ structure function, for instance, in
this framework is given by the sum of squares of form factors,
$F_{N\to R}(\vec{q\, }^2)$, describing the transitions from the nucleon
to excited states $R$,
\begin{eqnarray}
F_1(\nu,\vec{q\, }^2)
&\sim& \sum_R \left| F_{N\to R}(\vec{q\, }^2) \right|^2
	\delta(E_R-E_N-\nu)\ ,
\end{eqnarray}
where $E_N$ and $E_R$ are the energies of the ground state and excited
state, respectively.
In terms of photoabsorption cross sections (or $W$ boson absorption
cross sections for neutrino scattering), the $F_1$ structure function
is proportional to the sum $\sigma_{1/2} + \sigma_{3/2}$, with
$\sigma_{1/2 (3/2)}$ the cross section for total boson--nucleon
helicity 1/2 (3/2).
The spin-dependent $g_1$ structure function, on the other hand,
corresponds to the difference $\sigma_{1/2} - \sigma_{3/2}$.

Resonance excitation and deep inelastic scattering in general involve
both electric and magnetic multipoles.
Excitation in a given partial wave at $Q^2 = 0$ involves a complicated
mix of these.
However, as $Q^2$ grows one expects the magnetic multipole to dominate
over the electric, even by $Q^2 \sim 0.5$~GeV$^2$ in specific models
\cite{CGK,CG}.
Furthermore, recent phenomenological analyses of electromagnetic
excitations of negative parity resonances suggests that for the
prominent $D_{13}$ resonance the ratio of helicity-1/2 to helicity-3/2
amplitudes is consistent with zero beyond $Q^2 \approx 2$~GeV$^2$
\cite{BURKERT}, which corresponds to magnetic dominance.
This dominance of magnetic, or spin flip, interactions at large $Q^2$
for $N^*$ excitation matches the dominance of such spin flip in deep
inelastic scattering.
For instance, the polarization asymmetry $A_1 = g_1/F_1$ is positive
at large $Q^2$, whereas $A_1 < 0$ if electric interactions were
prominent \cite{FC74}.
Thus in the present analysis we assume that the interaction with the
quark is dominated by the magnetic coupling.
In this approximation the $F_1$ and $F_2$ structure functions are
simply related by the Callan-Gross relation, $F_2 = 2 x F_1$,
independent of the specific models we use for the structure functions
themselves.

The relative photoproduction strengths of the transitions from the
ground state to the {\bf 56}$^+$ and {\bf 70}$^-$ are summarized in
Table~I for the $F_1$ and $g_1$ structure functions of the proton
and neutron.
For generality, we separate the contributions from the symmetric and
antisymmetric components of the ground state nucleon wave function,
with strengths $\lambda$ and $\rho$, respectively.
The SU(6) limit corresponds to $\lambda = \rho$.
The coefficients in Table~I assume equal weights for the {\bf 56}$^+$
and {\bf 70}$^-$ multiplets \cite{CGK}.
Similarly, neutrino-induced transitions to excited states can be
evaluated \cite{CGNU}, and the relative strengths are displayed in
Table~II for the proton and neutron.
Because of charge conservation, only transitions to decuplet
(isospin-${3 \over 2}$) states from the proton are allowed.
(Note that the overall normalizations of the electromagnetic and
neutrino matrix elements in Tables~I and II are arbitrary.)

Summing over the full set of states in the {\bf 56}$^+$ and {\bf 70}$^-$
multiplets leads to definite predictions for neutron and proton structure
function ratios,
\begin{eqnarray}
R^{np} &=& { F_1^n \over F_1^p }\ ,		\\
R^\nu  &=& { F_1^{\nu p} \over F_1^{\nu n} }\ ,
\end{eqnarray}
and polarization asymmetries,
\begin{eqnarray}
A_1^N       &=& { g_1^N \over F_1^N}\ ,		\\
A_1^{\nu N} &=& { g_1^{\nu N} \over F_1^{\nu N}}\ ,
\end{eqnarray}
for $N=p$ or $n$.
In particular, for $\lambda=\rho$ one finds the classic SU(6)
quark-parton model results \cite{CLOSEBOOK}:
\begin{eqnarray}
\label{su6}
R^{np} &=& { 2 \over 3 }\ ,\ \ \ \
A_1^p\ =\ { 5 \over 9 }\ ,\ \ \ \
A_1^n\ =\ 0\ \ \ \ \ \ \ {\rm [SU(6)]}\ ,
\end{eqnarray}
for electromagnetic scattering, and
\begin{eqnarray}
\label{su6nu}
R^\nu &=& { 1 \over 2 }\ ,\ \ \ \
A_1^{\nu p}\ =\ -{ 1 \over 3 }\ ,\ \ \ \
A_1^{\nu n}\ =\ { 2 \over 3 }\ \ \ \ \ \ \ {\rm [SU(6)]}\ ,
\end{eqnarray}
for neutrino scattering, which correspond to $u=2d$ and
$\Delta u = -4 \Delta d$.
The quark level results are easily deduced by considering the wave
function of a proton in the SU(6) limit, polarized in the $+z$ direction
\cite{CLOSEBOOK}:
\begin{eqnarray}
\label{pwfn}
| p^\uparrow \rangle
&=& {1 \over \sqrt{2}}  | u^\uparrow (ud)_0 \rangle\
 +\ {1 \over \sqrt{18}} | u^\uparrow (ud)_1 \rangle\
 -\ {1 \over 3}         | u^\downarrow (ud)_1 \rangle\ \nonumber \\
& &
 -\ {1 \over 3}         | d^\uparrow (uu)_1 \rangle\
 -\ {\sqrt{2} \over 3}  | d^\downarrow (uu)_1 \rangle\ ,
\end{eqnarray}
where the subscript 0 or 1 denotes the total spin of the two-quark
component.
The neutron wave function is obtained from Eq.~(\ref{pwfn}) by
interchanging $u \leftrightarrow d$.
In this limit, apart from charge and flavor quantum numbers, the $u$ and
$d$ quarks in the proton are identical, and, in particular, have the same
$x$ distributions.
The relations between the structure functions and leading order parton
distributions are given in the Appendix.
The various structure function ratios in the SU(6) quark model are
listed in the first column of Table~III.

One should point out that these results arise in an ideal world of SU(6)
symmetry where the members of a {\bf 56}$^+$ or {\bf 70}$^-$ are each
degenerate, with common $Q^2$ dependent form factors.
Reality is not like that.
In the quark model the usual assignments of the excited states have the
nucleon and $P_{33}(1232)$ $\Delta$ isobar belonging to the quark
spin-${1\over 2}$ $^2${\bf 8} and quark spin-${3 \over 2}$ $^4${\bf 10}
representations of {\bf 56}$^+$, respectively, while for the odd parity
states the $^2${\bf 8} representation contains the states $S_{11}(1535)$
and
$D_{13}(1520)$, the $^4${\bf 8} contains the $S_{11}(1650)$,
$D_{13}(1700)$ and $D_{15}(1675)$, while the isospin-${3 \over 2}$ states
$S_{31}(1620)$ and $D_{33}(1700)$ belong to the $^2${\bf 10}
representation.
One purpose of this paper will be to investigate the systematics of such
SU(6) breaking which split energy levels, give different $Q^2$ dependence
to form factors, distort the $u$ and $d$ flavors and spin distributions,
and affect the $x \to 1$ behaviors via duality.

\section{Duality and SU(6) Breaking}

While the SU(6) predictions for the structure functions hold
approximately at $x \sim 1/3$, significant deviations are observed at
larger $x$.
Empirically, the $d$ quark distribution is observed to be much softer
than the $u$ for $x \agt 0.5$ \cite{CLOSEBOOK,WHITLOW,GOMEZ,NP}, leading
to $F_2^n/F_2^p \ll 2/3$ at large $x$.
Also, on the basis of helicity conservation \cite{FJ,GUNION}, one
expects that the proton and neutron polarization asymmetries, for both
electromagnetic and neutrino scattering, $A_1^N, A_1^{\nu N} \to 1$ as
$x \to 1$, in dramatic contrast to the SU(6) expectations, especially
for the neutron, where $A_1^n = 0$ and $A_1^{\nu n} = -1/3$.

In this Section we examine the conditions under which combinations of
resonances can reproduce, via quark-hadron duality, the behavior
of structure functions in the large-$x$ region where SU(6) breaking
effects are most prominent.
At the quark level, explicit SU(6) breaking mechanisms produce
different weightings of components of the initial state wave function,
Eq.~(\ref{pwfn}), which in turn induces different $x$ dependences for
the spin and flavor distributions.
On the other hand, at the hadronic level SU(6) breaking in the
$N \to N^*$ matrix elements leads to suppression of transitions to
specific resonances in the final state, while starting from a symmetric
SU(6) initial state wave function.
Thus if we admit breaking of the SU(6) symmetry, then for duality to be
manifest the pattern of symmetry breaking in the initial state has to
match that in the final state.

Note that for a fixed $W=M_R$ of a given resonance $R$, the resonance
peak moves to larger $x$ with increasing $Q^2$, since at the resonance
peak one has $x = x_R \equiv Q^2 / (M_R^2 - M^2 + Q^2)$.
At low $Q^2$, the prominent resonances are spread out in $x$ and a
necessary condition for duality involves integrating over a range of $x$
corresponding to $W \alt 2$~GeV.
At large $Q^2$ for fixed $x$ one has large $W$ and hence a dense
population of overlapping coherent resonance states.
In such a circumstance duality can become locally satisfied.
In turn this kinematics means that if a given resonance at $x \sim 1/3$
appears at relatively low $Q^2$, the $x \sim 1$ behavior of the
resonance contribution to the structure function will be determined by
the $N \to R$ transition form factor at large $Q^2$.

We shall look therefore for different $Q^2$ dependences in the
transition form factors to different spin-flavor multiplets, and study
their implications for $x \to 1$ in the sum.
Then we shall look at specific examples of resonances having these
particular correlations and identify experimental tests of the
hypothesis.

\subsection{Suppression of $\Delta$ states}

The most immediate breaking of the SU(6) duality could be achieved by
varying the overall strengths of the coefficients for the {\bf 56}$^+$
and {\bf 70}$^-$ multiplets as a whole.
However, since the cancellations of the $N \to N^*$ transitions for the
case of $g_1^n$ occur within each multiplet, a non-zero value of $A_1^n$
can only be achieved if SU(6) is broken {\em within} each multiplet
rather than {\em between} the multiplets.
Some intuition is needed therefore on sensible breaking patterns within
the supermultiplets.

Turning first to the {\bf 56}$^+$, empirical evidence suggests that at
high $Q^2$ the $N \to \Delta$ transition form factor is anomalously
suppressed relative to the elastic nucleon form factors \cite{DELTA,CM}.
This phenomenon has been attributed to spin-dependent forces between
quarks, such as from single gluon exchange \cite{DGG}, which split the
nucleon and $\Delta$ masses and necessarily break SU(6).
Removing the $^4${\bf 10}[{\bf 56}$^+$] from the $s$-channel sum causes
$R^{np}$ to fall (to $10/19 \approx 0.53$), as required
phenomenologically, and both $A_1^p$ and $A_1^n$ to increase (to 1
and 2/5, respectively) compared with the SU(6) values (see column 2
of Table~III.

Investigation of the coefficients in Tables~I and II, however, shows
that a suppression of the $\Delta$ alone is not consistent with
quark-hadron duality.
In particular, it gives rise to a $\Delta u/u$ ratio, extracted from
the electromagnetic structure functions (see Eq.~(\ref{Du_u}) in the
Appendix), which is greater than unity, thereby violating a partonic
interpretation of the structure functions.
Similarly, suppression of all decuplet contributions, namely the
$^4${\bf 10} in the {\bf 56}$^+$ and $^2${\bf 10} in the {\bf 70}$^-$
(column 3 of Table~III), still gives a value for the extracted
$\Delta u/u$ which exceeds unity.

The reason for the failure of duality here is that eliminating $\Delta$
states in the $s$-channel sum spoils the cancellation of exotic
exchanges in the $t$-channel, $\gamma \gamma \to N \bar{N}$.
Non-exotic {\bf 1} and {\bf 35} SU(6) representations correspond to
$q\bar{q}$, thus in the $t$-channel these appear as
$\gamma \gamma \to q\bar{q}$; when such a diagram is viewed in the
$s$-channel one sees that in effect it can map onto handbag or leading
twist topologies, enabling a partonic interpretation, as shown in
Fig.~1(a).
Exotic exchanges, such as {\bf 405}, require $qq\bar{q}\bar{q}$ in the
$t$-channel, and map onto higher twist contributions, such as in
Fig.~1(b).
These are incompatible with single parton probability interpretations in
principle, with the specific $\Delta u/u > 1$ result illustrating this.

Moreover, the results for the $\Delta u/u$ and $\Delta d/d$ ratios
extracted from the electromagnetic observables, namely $\Delta u/u=23/21$
and $\Delta d/d=-1/3$, do not agree with those obtained from the neutrino
polarization asymmetries $A_1^{\nu N}$ (column 3 of Table~III.
In addition, for both of these scenarios the ratio $d/u$ extracted from
$R^{np}$ does not match that obtained from $R^\nu$.
These are all consequences of the presence of $t$-channel exotics in
such scenarios, and further underscore the inconsistency of duality with
suppression of $\Delta$ states alone.

\subsection{Spin 3/2 suppression}

If the characteristic $Q^2$ dependence for $\Delta$ excitation is indeed
due to spin dependence, then it may be that this is a phenomenon realized
by all $S=3/2$ quark couplings, namely
$^4${\bf 10}[{\bf 56}$^+$] and $^4${\bf 8}[{\bf 70}$^-$].
An immediate observation in this scenario from Tables~I and II is that
each of the contributions corresponding to (the surviving) quark spin
$S=1/2$ configurations has equal strength for $g_1$ and $F_1$, which
automatically gives unity for the polarization asymmetries $A_1$.
This simply follows from the (high $Q^2$) approximation that only
magnetic couplings to quarks contribute, so that only $S=3/2$
configurations allow non-zero $\sigma_{3/2}$ cross sections
(we shall return to this later).

More generally, one can observe that duality is satisfied by summing
over the individual $S=1/2$ and $S=3/2$ contributions separately,
$S_{1/2}\equiv$ $^2${\bf 8}[{\bf 56}$^+$] + $^2${\bf 8}[{\bf 70}$^-$]
		+  $^2${\bf 8}[{\bf 70}$^-$],
and
$S_{3/2}\equiv$ $^4${\bf 10}[{\bf 56}$^+$] + $^4${\bf 8}[{\bf 70}$^-$].
If the relative contributions of the $S_{1/2}$ and $S_{3/2}$ channels
are weighted by $\cos^2\theta_s$ and $\sin^2\theta_s$, respectively,
then the ratio of unpolarized neutron to proton structure functions
can be written as:
\begin{eqnarray}
\label{np_s}
R^{np}
&=& { 6 ( 1 + \sin^2\theta_s) \over 19 - 11 \sin^2\theta_s }\ ,
\end{eqnarray}
and the polarization asymmetries become:
\begin{eqnarray}
\label{a1p_s}
A_1^p &=& { 19 - 23 \sin^2\theta_s \over 19 - 11 \sin^2\theta_s }\ , \\
\label{a1n_s}
A_1^n &=& { 1 - 2 \sin^2\theta_s \over 1 + \sin^2\theta_s }\ .
\end{eqnarray}
The dependence on the mixing angle $\theta_s$ of these ratios is
illustrated in Fig.~2 (dashed curves).
The SU(6) symmetric limit, Eq.~(\ref{su6}), is reproduced when
$\theta_s=\pi/4$, as indicated in Fig.~2.
As $\theta_s \to 0$, corresponding to $S_{1/2}$ dominance, the neutron
to proton ratio decreases, and both the polarization asymmetries
approach their maximal values,
\begin{eqnarray}
\label{s12}
R^{np} &=& { 6 \over 19 }\ ,\ \ \ \
A_1^p\ =\ 1\ ,\ \ \ \
A_1^n\ =\ 1\ \ \ \ \ \ \ [\theta_s=0]\ .
\end{eqnarray}
In the other extreme limit as $\theta_s \to \pi/2$, the polarization
asymmetries approach $-1$, while $R^{np} \to 3/2$.
Neither of these scenarios are supported phenomenologically, as we
shall discuss below, and the physical region appears to correspond to
$0 \alt \theta_s \alt 9\pi/32$.

In analogy with Eqs.~(\ref{np_s})--(\ref{a1n_s}), the ratio of the
unpolarized proton and neutron structure functions for neutrino
scattering is:
\begin{eqnarray}
\label{fnu_s}
R^\nu
&=& { 1 + 7 \sin^2\theta_s \over 14 - 10 \sin^2\theta_s }\ ,
\end{eqnarray}
and the neutrino polarization asymmetries:
\begin{eqnarray}
\label{anup_s}
A_1^{\nu p} &=& { 1 - 5 \sin^2\theta_s \over 1 + 7 \sin^2\theta_s }\ , \\
\label{anun_s}
A_1^{\nu p} &=& { 7 - 8 \sin^2\theta_s \over 7 - 5 \sin^2\theta_s }\ .
\end{eqnarray}
The dependence on the angle $\theta_s$ for the neutrino observables is
shown in Fig.~3 (dashed curves).
The trends of the ratios are similar to those of the electromagnetic
ratios in Fig.~2 (with the neutron and proton reversed).
Once again the SU(6) symmetric limit, Eq.~(\ref{su6nu}), is reproduced
when $\theta_s=\pi/4$.
The phenomenologically favored scenario in which $S_{3/2}$ contributions
are suppressed in the limit $x \to 1$ gives rise to:
\begin{eqnarray}
\label{s12nu}
R^\nu &=& { 1 \over 14 }\ ,\ \ \ \
A_1^{\nu p}\ =\ 1\ ,\ \ \ \
A_1^{\nu n}\ =\ 1\ \ \ \ \ \ \ [\theta_s=0]\ .
\end{eqnarray}
{}From the relations between the structure functions and parton
distributions in the Appendix one can verify that the results for $d/u$
extracted from $R^{np}$ are consistent with those from $R^\nu$
(Eqs.~(\ref{du}) and (\ref{du_nu})), and those for $\Delta q/q$
extracted from $A_1^N$ consistent with those from $A_1^{\nu N}$
(Eqs.~(\ref{Du_u})--(\ref{Dd_d})) and
 Eqs.~(\ref{A1nup_qpm})--(\ref{A1nun_qpm})).

The dependence of the structure function ratios in
Eqs.~(\ref{np_s})--(\ref{a1n_s}) and Eqs.~(\ref{fnu_s})--(\ref{anun_s})
on one parameter, $\theta_s$, means that the SU(6) breaking scenario
with $S_{3/2}$ suppression can be tested by simultaneously fitting the
$n/p$ ratios and the polarization asymmetries.
In general, data on unpolarized structure functions are more abundant,
especially at high $x$, than on spin dependent structure functions, so
it is more practical to fit the $x$ dependence of $\theta_s(x)$ to the
existing data on unpolarized $n/p$ ratios, which can then be used to
predict the polarization asymmetries.

Unfortunately, data on $F_1$ neutrino structure functions at
$x \agt 0.4$--0.5 are essentially non-existent, and there have been
no experiments at all to measure spin-dependent structure functions in
neutrino scattering.
The most precise data on the electromagnetic neutron to proton ratio
$R^{np}$ comes from SLAC experiments \cite{WHITLOW,GOMEZ}.
The absence of free neutron targets has meant that neutron structure
information has had to be inferred from inclusive deuteron and proton
structure functions.
Because of uncertainties in the treatment of nuclear corrections in the
deuteron at large $x$, however, which is more sensitive to the high
momentum components of the deuteron wave function, the results beyond
$x \sim 0.6$ are somewhat model dependent \cite{NP}, as indicated in
Fig.~4.
The difference between the two sets of points is representative of the
theoretical uncertainty in the extraction.
In particular, the lower set of points corresponds to an analysis which
accounts for Fermi motion in the deuteron \cite{FS}, while the upper set
of points includes Fermi motion and binding effects \cite{NP} (see also
Ref.~\cite{NUCLEAR}).
A fit to the weighted average of the extrema of the two sets of data
points, constrained to approach $R^{np}=6/19$ as $x \to 1$, is indicated
by the dashed curve (a polynomial of degree two is used to fit the
$x$ dependence of $\theta_s(x)$ in Eq.~(\ref{np_s})).
The fit is clearly compatible with the current data on $R^{np}$, but
could be further constrained by more accurate data at large $x$.
Several proposals for obtaining the neutron to proton ratio at large $x$
with reduced nuclear uncertainties are discussed in Refs.~\cite{A3,BONUS}.

Using the mixing angle $\theta_s(x)$ fitted to $R^{np}$, the resulting
polarization asymmetries for the proton and neutron are shown in Figs.~5
and 6, respectively, compared with a compilation of large-$x$ data from
SLAC \cite{SLAC}, SMC \cite{SMC} and HERMES \cite{HERMES}.
The predicted $x$ dependence of both $A_1^p$ and $A_1^n$ in the $S_{3/2}$
suppression scenario is relatively strong; the SU(6) symmetric results
which describe the data at $x \sim 1/3$ rapidly give way to the
broken SU(6) predictions as $x \to 1$.
Within the current experimental errors, the $S_{3/2}$ suppression model
is consistent with the $x$ dependence of both the $R^{np}$ ratio and the
polarization asymmetries.

Using the neutrino ratios $R^\nu$, $A_1^{\nu p}$ and $A_1^{\nu n}$,
the individual quark flavor and spin distribution ratios can be
determined (or equivalently, extracted from the electromagnetic ratios
as discussed in the Appendix).
The unpolarized $d/u$ ratio in the $S_{1/2}$ dominance scenario is shown
in Fig.~7 (dashed), and the spin-flavor ratios $\Delta u/u$ and
$\Delta d/d$ are illustrated in Figs.~8 and 9, respectively.

\subsection{Helicity 3/2 suppression}

The above discussion has demonstrated how duality between the parton model
and a sum over low-lying resonances can arise on the basis of classifying
transitions to excited states according to the total spin of the quarks,
with either equal weighting of $S_{1/2}$ and $S_{3/2}$ components in the
case of SU(6) symmetry, or suppression of the latter at large $x$.
According to duality, structure functions at large $x$ are determined
by the behavior of transition form factors at high $Q^2$; hence one may
expect that at large enough $Q^2$ these would be constrained by
perturbative QCD.
In particular, at high $Q^2$ perturbative arguments suggest that the
interaction of the photon (or $W$ boson) should be predominantly with
quarks with the same helicity as the nucleon \cite{FJ,GUNION}.
Since the photon ($W$ boson) scattering from a massless quark conserves
helicity, the $\sigma_{3/2}$ cross section would be expected to be
suppressed relative to the $\sigma_{1/2}$ \cite{CLOSEBOOK}.
The question then arises whether duality can exist between parton
distributions at large $x$ and resonance transitions classified according
to quark {\em helicity} rather than spin.

In general, if the relative strengths of the $\sigma_{1/2}$ and
$\sigma_{3/2}$ contributions to the cross section are weighted by
$\cos^2\theta_h$ and $\sin^2\theta_h$, respectively, then from Table~I
the ratio of the neutron to proton $F_1$ structure functions can be
written as:
\begin{eqnarray}
R^{np}
&=& { 3 \over 7 - 5 \sin^2\theta_h }\ ,
\end{eqnarray}
while the proton and neutron polarization asymmetries become:
\begin{eqnarray}
A_1^p &=& { 7 - 9 \sin^2\theta_h \over 7 - 5 \sin^2\theta_h }\ , \\
A_1^n &=& 1 - 2 \sin^2\theta_h\ .
\end{eqnarray}
Similarly for neutrino scattering, one has:
\begin{eqnarray}
R^\nu
&=& { 1 + \sin^2\theta_h \over 5 - 4 \sin^2\theta_h }
\end{eqnarray}
for the unpolarized structure functions, and
\begin{eqnarray}
A_1^{\nu p} &=& { 1 - 3 \sin^2\theta_h \over 1 + \sin^2\theta_h }\ , \\
A_1^{\nu n} &=& { 5 - 6 \sin^2\theta_h \over 5 - 4 \sin^2\theta_h }\
\end{eqnarray}
for neutrino induced polarization asymmetries.
The dependence of these ratios on the mixing angle $\theta_h$ is
illustrated in Figs.~2 and 3 (solid curves).
For $\theta_h=\pi/4$ the SU(6) results in Eqs.~(\ref{su6}) and
(\ref{su6nu}) are once again recovered.
In the phenomenologically favored region of $0 \leq \theta_h \leq \pi/4$
the predictions for $A_1^p$ and for $A_1^{\nu n}$ are very similar to
those derived on the basis of quark spin, which reflects the fact that
the ratios $\Delta u/u$ are predicted to be similar in both cases.
Both the $\sigma_{3/2}$ and $S_{3/2}$ suppression scenarios give rise to
the same predictions for $A_1^n$ in the $\theta \to 0$ limit, although
the approach to the maximum values is faster in the case of
$\sigma_{3/2}$ suppression.
For the unpolarized ratios, $\sigma_{3/2}$ suppression gives rise to
larger values of $R^{np}$ and $R^\nu$ than for $S_{3/2}$ suppression.
This is also evident from the modified transition strengths for $F_1$
and $g_1$ displayed in Tables~IV and V for the case of $\sigma_{1/2}$
dominance at large $x$.
Summing up the coefficients for the neutron and proton, one has in the
limit $x \to 1$:
\begin{eqnarray}
\label{sig12}
R^{np} &=& { 3 \over 7 }\ ,\ \ \ \
A_1^p\ =\ 1\ ,\ \ \ \
A_1^n\ =\ 1\ \ \ \ \ \ \ [\theta_h=0]\ ,
\end{eqnarray}
for the electromagnetic ratios, and
\begin{eqnarray}
\label{sig12nu}
R^\nu &=& { 1 \over 5 }\ ,\ \ \ \
A_1^{\nu p}\ =\ 1\ ,\ \ \ \
A_1^{\nu n}\ =\ 1\ \ \ \ \ \ \ [\theta_h=0]\ ,
\end{eqnarray}
for neutrino scattering.

Fitting the $x$ dependence of the mixing angle $\theta_h(x)$ to the
$R^{np}$ data with the above $x \to 1$ constraint (Fig.~4), the
resulting predictions for $A_1^{p,n}$ are shown in Figs.~5 and 6,
respectively.
Compared with the $S_{1/2}$ dominance scenario, the $\sigma_{1/2}$
dominance model predicts a faster approach to the asymptotic limits.
The values for the ratios in Eqs.~(\ref{sig12}) and (\ref{sig12nu})
correspond exactly to those calculated at the quark level on the basis
of perturbative QCD counting rules \cite{FJ,GUNION}.
There, the deep inelastic scattering at $x \sim 1$ requires the
exchange in the initial state of two hard gluons, which preferentially
enhances those configurations in the nucleon wave function in which the
spectator quarks have zero helicity.
The structure function at large $x$ is then determined by components of
the nucleon wave function in which the helicity of the interacting
quark matches that of the nucleon.
For an initial state SU(6) wave function, Eq.~(\ref{pwfn}), suppression
of the helicity anti-aligned configurations leads to the unpolarized
ratio $d/u=1/5$, and the polarization ratio $\Delta q/q = 1$ for all
quark flavors.
Using the relations in the Appendix between the structure functions and
the leading order quark distributions, one can verify the equivalence
of the parton- and hadron-level results via quark-hadron duality.

The resulting quark-level ratios are shown in Fig.~7 for the $d/u$ ratio,
and in Figs.~8 and 9 for the $\Delta u/u$ and $\Delta d/d$ ratios,
respectively.
While the behavior of the $\Delta u/u$ ratio is similar in both the
$S_{1/2}$ and $\sigma_{1/2}$ dominance models, the predicted
$\Delta d/d$ ratio has a more rapid approach to unity for the latter
case.

\subsection{Symmetric wave function suppression}

In SU(3)$\times$SU(2) the relevant multiplets are the
spin-${1 \over 2}$ $^2${\bf 8} and $^2${\bf 10}, and
spin-${3 \over 2}$ $^4${\bf 8} and $^4${\bf 10}.
In SU(6) the $^2${\bf 10} and $^4${\bf 8} multiplets are in the
{\bf 70}$^-$ representation, and the $^4${\bf 10} unambiguously in the
{\bf 56}$^+$ representation.
However, the $^2${\bf 8} occur in both the {\bf 56}$^+$ and {\bf 70}$^-$.
In general, for the $^2${\bf 8} states one can write the nucleon wave
function in terms of symmetric and antisymmetric components,
\begin{eqnarray}
\label{rholam}
| N \rangle &=& \cos\theta_w | \psi_\rho \rangle\
	     +\ \sin\theta_w | \psi_\lambda \rangle\ ,
\end{eqnarray}
where $\psi = \varphi \otimes \chi$ is a product of the flavor ($\varphi$)
and spin ($\chi$) wave functions, and $\lambda$ and $\rho$ denote the
symmetric and antisymmetric combinations, respectively \cite{CLOSEBOOK}.
In the SU(6) limit one has an equal admixture of both $\rho$ and
$\lambda$ type contributions, $\theta_w = \pi/4$, and the symmetric
wave function of Eq.~(\ref{pwfn}) is recovered.

The SU(6) symmetry can be broken if the mixing angle
$\theta_w \not= \pi/4$.
In particular, if the mass difference between the nucleon and $\Delta$
is attributed to spin dependent forces, the energy associated with the
symmetric part of the wave function will be larger than that of the
antisymmetric component.
A suppression of the symmetric $| \psi_\lambda \rangle$ configuration at
large $x$ will then give rise to a suppressed $d$ quark distribution
relative to $u$, $d/u \to 0$, which in turn leads to the extreme limits
for the $R^{np}$ and $R^\nu$ ratio allowed by the quark parton model,
$R^{np} \to 1/4$ and $R^\nu \to 0$ \cite{QUARTER}.
It also leads to the proton and neutron polarization asymmetries
becoming unity as $x \to 1$ \cite{FC74}.
At the parton level, this pattern of suppression can be realized, for
instance, with a spin-dependent hyperfine interaction between quarks,
$\vec S_i \cdot \vec S_j$, which modifies the spin-0 and spin-1
components of the nucleon wave function and leads to a softening of
the $d$ quark distribution relative to the $u$ at large $x$
(see Ref.~\cite{QUARTER} for details).

This scenario is also consistent with the absence of exotics in the
$t$-channel.
This can be demonstrated by examining the pattern of suppressions in the
structure function calculated, via quark-hadron duality, from the sum
over resonances in the final state.
In this case, the symmetric components of the states in the
{\bf 56}$^+$ and {\bf 70}$^-$ multiplets are suppressed relative to the
antisymmetric, and the modified relative transition strengths are given
in Table~I with $\lambda \to 0$.
In particular, since transitions to the (symmetric) $S=3/2$ or decuplet
states ($^4${\bf 8}, $^4${\bf 10} and $^2${\bf 10}) can only proceed
through the symmetric ``$\lambda$'' component of the ground state wave
function, the ``$\rho$'' components will only excite the nucleon to
$^2${\bf 8} states.
If the $\lambda$ wave function is suppressed, only transitions to
$^2${\bf 8} states will be allowed.
Summing over all channels leads to an unpolarized neutron to proton
ratio in terms of the mixing angle $\theta_w$ given by:
\begin{eqnarray}
R^{np}
&=& { 1 + 2 \sin^2\theta_w \over 4 - 2 \sin^2\theta_w }\ ,
\end{eqnarray}
with polarization asymmetries given by:
\begin{eqnarray}
A_1^p &=& { 6 - 7 \sin^2\theta_w \over 6 - 3 \sin^2\theta_w }\ , \\
A_1^n &=& { 1 - 2 \sin^2\theta_w \over 1 + 2 \sin^2\theta_w }\ .
\end{eqnarray}
The dependence on $\theta_w$ is shown in Fig.~2.
In the limit of $\rho$ dominance at $x \to 1$, one recovers the ratios:
\begin{eqnarray}
\label{rho}
R^{np} &=& { 1 \over 4 }\ ,\ \ \ \
A_1^p\ =\ 1\ ,\ \ \ \
A_1^n\ =\ 1\ \ \ \ \ \ \ [\theta_w=0]\ .
\end{eqnarray}
Fitting $\theta_w$ to the $x$ dependence of $R^{np}$ in Fig.~4 with the
above constraints (dot-dashed), the resulting $x$ dependence of the
polarization asymmetries $A_1^p$ and $A_1^n$ are shown in Figs.~5 and 6
(dot-dashed).
The approach to the asymptotic values for the polarization asymmetries
is less rapid than for the $\sigma_{1/2}$ or $S_{1/2}$ dominance
scenarios.

Similarly, for neutrino scattering, one has:
\begin{eqnarray}
R^\nu
&=& { 2 \sin^2\theta_w \over 3 - 2 \sin^2\theta_w }\ ,
\end{eqnarray}
and
\begin{eqnarray}
A_1^{\nu p} &=& - { 1 \over 3 }\ ,	\\
A_1^{\nu n} &=& { 9 - 10 \sin^2\theta_w \over 9 - 6 \sin^2\theta_w }\ ,
\end{eqnarray}
for neutrino induced polarization asymmetries.
Note that the neutrino--proton polarization asymmetry remains negative,
as in SU(6), and is independent of the mixing angle.
The dependence on $\theta_w$ of the ratios is illustrated in
Figs.~2 and 3 (dot-dashed curves), where in the limit $\theta_w \to 0$
($x \to 1$) one has:
\begin{eqnarray}
R^\nu &=& { 0 }\ ,\ \ \ \
A_1^{\nu p}\ =\ - {1 \over 3}\ ,\ \ \ \
A_1^{\nu n}\ =\ 1\ \ \ \ \ \ \ [\theta_w=0]\ .
\end{eqnarray}
The ratios of the associated quark densities are given in
Figs.~7--9 for $d/u$, $\Delta u/u$ and $\Delta d/d$, respectively.
Because the neutron asymmetry $A_1^{\nu n}$ is negative, the predicted
$\Delta d/d$ ratio has qualitatively different behavior in the
$\lambda$ suppression scenario than in the other two SU(6) broken
models.
It would clearly be of considerable interest to test the behavior of
$\Delta d/d$ experimentally, for instance in semi-inclusive deep
inelastic scattering by tagging pions.

\section{Implications for Low Lying Resonances}

If the suppression of specific spin-flavor correlations, as required to
fit the $x \to 1$ behavior of structure functions, is a property of
spin-dependent inter-quark forces, then they should affect specific
resonances that share these properties.
In this Section we identify some examples and propose measurements
that can test the veracity of the various scenarios discussed in
Section~III.

\subsection{Suppression of $^4${\bf 10} states}

If the suppression of the $P_{33}(1232)$ $\Delta$ isobar at large $Q^2$
is characteristic of $^4${\bf 10} and $^4${\bf 8} states, then a
careful study of electroproduction of the $L=2$ {\bf 56}$^+$ states
$P_{31}(1930)$, $P_{33}(1920)$, $F_{35}(1905)$ and $F_{37}(1950)$ may
reveal $S_{3/2}$ suppression as the appropriate physical mechanism
responsible for symmetry breaking in structure functions at large $x$.
Transitions to each of these states, in the absence of configuration
mixing, should die relatively faster with $Q^2$ than for the
$^2${\bf 8} and $^2${\bf 10} resonances.
This should be particularly so for the $F_{37}(1950)$, where mixing
should be minimal, although one must ensure to have gone past the high
angular momentum threshold that may cause the form factors for high spin
states to remain large in the small $Q^2$ region.

A possible way to normalize the production, and cancel out such
threshold enhancements, will be to compare the relative strengths of
these $^4${\bf 10} and their partner $^2${\bf 8}[{\bf 56}$^+$] states.
Thus measurement of the $Q^2$ dependence of ratios such as
\[
F_{35}(1905)/ F_{15}(1680) ; P_{33}(1920)/P_{13}(1720)
\]
would be crucial in testing this scenario.

\subsection{Suppression of $^4${\bf 8} states}

In general, mixing is expected between the $^4${\bf 8} and $^2${\bf 8}
states with the same $J^P$.
For example, the physical $S_{11}(1550)$ and $S_{11}(1650)$ states are
superpositions of $^2${\bf 8} and $^4${\bf 8} components:
\begin{eqnarray}
S_{11}^a(1535)
&=& \cos\theta |^2 8 \rangle + \sin\theta | ^4 8 \rangle\ , \\
S_{11}^b(1650)
&=& \sin\theta |^2 8 \rangle - \cos\theta | ^4 8 \rangle\ ,
\end{eqnarray}
and similarly for the $D_{13}(1520)$ and $D_{13}(1700)$ states.
{}From protons one then expects:
\begin{eqnarray}
{\sigma (\gamma^* p \to S_{11}^a) \over
 \sigma (\gamma^* p \to S_{11}^b)}
&\sim& \cot^2\theta\ ,
\end{eqnarray}
which will be true for all $Q^2$, as the $^4${\bf 8} component is not
excited.
{}From neutron targets, however, both components are excited at low
$Q^2$, whereas the $^4${\bf 8} is suppressed at large $Q^2$.
Hence at small $Q^2$ one has:
\begin{eqnarray}
{\sigma (\gamma^* n \to S_{11}^a) \over
 \sigma (\gamma^* n \to S_{11}^b)}
&\sim& f(\theta)\ ,
\end{eqnarray}
where $f(\theta)$ is a function of the mixing angle and of the relative
strengths of the $^2${\bf 8} and $^4${\bf 8} photocouplings.
However, at large $Q^2$ only the $^2${\bf 8} is predicted to survive
(thus in effect the Moorhouse selection rule \cite{MOORHOUSE} will hold
for neutrons too when $Q^2 \to \infty$), in which case
\begin{eqnarray}
\left.
{\sigma (\gamma^* n \to S_{11}^a) \over
 \sigma (\gamma^* n \to S_{11}^b)}
\right|_{Q^2 \to \infty}
&\to& \cot^2\theta \equiv
\left.
{\sigma (\gamma^* p \to S_{11}^a) \over
 \sigma (\gamma^* p \to S_{11}^b)}
\right|_{{\rm all}\ Q^2}\ .
\end{eqnarray}
As this behavior is predicted to be common for $p$ and $n$ targets, it
should therefore hold true for the deuteron.
The $D_{15}(1690)$ is a pure $^4${\bf 8} state and so provides a clean
test of the fast $Q^2$ suppression in electroproduction from neutron
targets.

\subsection{Suppression of $\sigma_{3/2}$}

The suppression of helicity-${3 \over 2}$ contributions allows
transitions to the $\Delta$ to survive, as well as excitations to the
$^4${\bf 8} states from the neutron (those from the proton vanish
because of the Moorhouse selection rule \cite{MOORHOUSE}).
At $Q^2 = 0$, the $\Delta$ excitation is pure M1, which equates with
$\sigma_{3/2} = 3 \sigma_{1/2}$, and leads to the polarization
asymmetry $A_1^N = -1/2$.
At large $Q^2$ the survival of the $\Delta$, in the $\sigma_{3/2}$
channel, corresponds to the E2 excitation becoming comparable to the M1.

Electroproduction of the $S_{11}$, $D_{13}$ and $D_{15}$ resonances
from neutrons will change from $A_1^N = -1/2$ to $A_1^N = 1$.
This should remain true for the $D_{15}$, but can be obscured by mixing
with $^2${\bf 8} for the $S_{11}$ and $D_{13}$.
Configuration mixing between the $^2${\bf 8} and $^4${\bf 8} states
(with mixing angle $\sim 30^\circ$) does allow a relatively strong
transition to the $S_{11}(1650)$.
Data from CLAS at Jefferson Lab \cite{BURKERT} suggest that, within the
single quark transition model \cite{SQTM}, the strength of the
$S_{11}(1650)$ transition is about half of that to the $S_{11}(1535)$.
The mixing angle between the $^4${\bf 8} and $^2${\bf 8} states with
$J^P={3\over 2}^-$ is much smaller ($\sim 6^\circ$), so that
transitions to the $D_{13}(1700)$ will be weakly excited from the
proton.
The strengths for the other $^4${\bf 8} states are known only at
$Q^2=0$, so that data on these transition form factors at
$Q^2 \sim 1$--2~GeV$^2$ would be valuable in establishing the extent
of any suppression.

\subsection{Suppression of $\psi_\lambda$ wave function}

The consequences for $N^*$s in this scenario are quite extensive.
Namely, transitions to $^4${\bf 10}, $^4${\bf 8} and $^2${\bf 10}
states are all suppressed, and only transitions to $^2${\bf 8} are
allowed.
While the transitions for the proton to $^2${\bf 8} are unchanged
compared with the SU(6) case, for neutron the elastic transition is
reduced by $\sim 50\%$, and the transition to the {\bf 70}$^-$ enhanced
by $\sim 50\%$.

Another prediction of $\lambda$ wave function suppression is identical
production rates in both the {\bf 56}$^+$ and {\bf 70}$^-$ channels,
for electron and neutrino scattering.
For the latter, essentially no empirical information exists, however,
neutrino structure functions in the resonance region may be accessible
in the future at a high-intensity neutrino beam facility \cite{NUMI}.
In particular, since neutrinos can excite protons {\em only} to decuplet
states, this may provide a valuable test of the $\lambda$-suppression
mechanism, and of the isospin dependence of the $N \to N^*$ transitions.

\section{Conclusion}

In this analysis we have performed a first detailed study of the
conditions under which SU(6) symmetry breaking in the quark model can
yield consistent results for structure function ratios in the context
of quark-hadron duality.
Several self-consistent SU(6) breaking scenarios have been identified,
involving the suppression of transitions to states in the lowest even
and odd parity multiplets with quark spin $S=3/2$, to states with
helicity ${3 \over 2}$, and to states which couple only through
symmetric components of the wave function, $\psi_\lambda$.

The implications of the various symmetry breaking scenarios on the $x$
dependence of structure function ratios have been quantified, which can
be tested in future experimental studies.
In particular, fitting to the available data on the unpolarized neutron
to proton ratio $R^{np}$ allows one to make predictions for the large
$x$ behavior of polarization asymmetries $A_1^N$.
Experiments proposed at an energy-upgraded Jefferson Lab should enable
the $R^{np}$ ratio to be reliably determined up to $x \sim 0.85$
\cite{A3,BONUS}.
For the polarization asymmetries there is existing evidence that
$A_1^p > 5/9$ at $x \agt 0.6$, and recent data on $A_1^n$ from Hall~A
at Jefferson Lab \cite{A1N,XIAOCHAO} give the first hint of a rise
above zero at $x \sim 0.6$.
High precision data on $A_1^p$ or $A_1^n$ at large $x$ would help
constrain also the unpolarized $n/p$ ratio, and allow a simultaneous
test of the duality relations.

Measurement of the neutrino structure function ratios, on the other
hand, is more challenging due to the low rates at large $x$, and the
need for large volume (typically iron) targets, which is particularly
problematic for the spin-dependent observables.
The prospect of high intensity neutrino beams at Fermilab to measure
structure functions in the resonance region \cite{NUMI} offers a
valuable complement to the study of duality and resonance transitions.
A parallel avenue towards determining the spin-flavor asymmetries
such as $\Delta u/u$ and $\Delta d/d$, which is particularly sensitive
to different SU(6) breaking assumptions, could be provided through
a program of semi-inclusive scattering tagging fast pions in the
current fragmentation region.

A quantitative description of transition form factors in the quark
model at moderate $Q^2$ must involve both longitudinal and transverse
response, electric and magnetic couplings, as well as hyperfine
interactions which explicitly break SU(6) symmetry.
On the other hand, most of these complications do not affect the main
elements of duality, and can obfuscate the basic principles which drive
the quark-hadron transition.
For reasons of clarity, in the present analysis we have considered only
magnetic transitions, which are expected to dominate at high $Q^2$.
This assumption leads, for instance, to the electromagnetic neutron to
proton ratio $R^{np}=4/9$ for the case of elastic scattering, which is
equal to the squared ratio of the neutron to proton magnetic moments in
SU(6) \cite{BG,EL}.
Electric transitions would give a ratio $R^{np}=0$.
Electric couplings will also modify the coefficients in Tables~I and II
for the other transitions \cite{COT}.
Although electric couplings will play a role at low $Q^2$, for the
behavior of structure functions at large $x$ one expects magnetic
couplings to dominate the transition form factors at high $Q^2$.

In future we shall extend this work to the longitudinal structure
function, which will necessitate inclusion of electric couplings.
Questions about the role of higher excitations, such as in the $N=2$,
$L=2$ band, will also be important to elucidate in more refined
analyses.
There are a number of states with mass $W \alt 1.8$~GeV which belong
to higher multiplets, such as the $F_{15}(1680)$, which is believed to
play an important role in the third resonance region.
In addition, it will be interesting to ascertain the role played by the
the $P_{11}(1440)$ Roper resonance in duality, which may shed some
light on the long-standing question about its internal structure
\cite{LATTICE}.

\acknowledgements

We would like to thank V.~D.~Burkert for helpful discussions.
This work was supported by the U.S. Department of Energy contract
\mbox{DE-AC05-84ER40150}, under which the Southeastern Universities
Research Association (SURA) operates the Thomas Jefferson National
Accelerator Facility (Jefferson Lab),
and by grants from the Particle Physics and Astronomy Research Council,
and the EU program ``Eurudice" HPRN-CT-2002-0031.

\section*{Appendix: Parton model and duality relations}

Here we summarize the quark parton model relations between
electromagnetic and neutrino structure functions and leading order
parton distributions.
The spin-averaged and spin-dependent $F_1$ and $g_1$ structure functions
are expressed in terms of a sum and difference of helicity cross sections,
%
\begin{eqnarray}
F_1 &\sim& \sigma_{1/2} + \sigma_{3/2}\ ,	\\
g_1 &\sim& \sigma_{1/2} - \sigma_{3/2}\ ,
\end{eqnarray}
%
where $\sigma_{1/2(3/2)}$ is the cross sections corresponding to total
boson--nucleon helicity 1/2 (3/2).

In the parton model the structure functions for charged lepton scattering
can be expressed (at leading order) in terms of quark distribution
functions,
\begin{eqnarray}
F_1(x) &=& {1 \over 2} \sum_q e_q^2\ q(x)\ ,		\\
g_1(x) &=& {1 \over 2} \sum_q e_q^2\ \Delta q(x)\ ,
\end{eqnarray}
where $q = q^\uparrow + q^\downarrow$
and $\Delta q = q^\uparrow - q^\downarrow$.
Inverting these, one can similarly extract leading order quark
distributions from the measured structure functions.
For instance, the $d/u$ quark distribution ratio can be determined from:
\begin{eqnarray}
\label{du}
{ d \over u } &=& { 4 R^{np} - 1 \over 4 - R^{np} }\ ,
\end{eqnarray}
where $R^{np}=F_1^n/F_1^p$, while the spin dependent flavor ratios for
the $u$ and $d$ quarks are obtained from the polarization asymmetries and
the $d/u$ ratio in Eq.~(\ref{du}) \cite{XIAOCHAO},
\begin{eqnarray}
\label{Du_u}
{ \Delta u \over u }
&=& { 4 \over 15 } A_1^p \left( 4 +   { d\over u} \right)\
 -\ { 1 \over 15 } A_1^n \left( 1 + 4 { d\over u} \right)\ ,	\\
\label{Dd_d}
{ \Delta d \over d }
&=&-{ 1 \over 15 } A_1^p \left( 1 + 4 { u\over d} \right)\
 +\ { 4 \over 15 } A_1^n \left( 4 +   { u\over d} \right)\ ,
\end{eqnarray}
where
\begin{eqnarray}
A_1^p &=& { 4 \Delta u + \Delta d \over 4 u + d }\ ,	\\
A_1^n &=& { \Delta u + 4 \Delta d \over u + 4 d }\ .
\end{eqnarray}
Note that if $A_1^p = A_1^n \equiv A_1^N$, then
$\Delta u/u = \Delta d/d = A_1^N$, independent of the value of $d/u$.

For neutrino scattering one has:
\begin{eqnarray}
F_1^\nu(x) &=& \sum_q g_q^2\ q(x)\ ,		\\
g_1^\nu(x) &=& \sum_q g_q^2\ \Delta q(x)\ ,
\end{eqnarray}
where for protons $g_q^2 = 1$ for $q=d, \bar u, \cdots$
and 0 for $q=u, \bar d, \cdots$, and vice versa for neutrons.
At large $x$ therefore $F_1^{\nu p}$, $g_1^{\nu p}$ directly probe the
$d$ quark distributions, while $F_1^{\nu n}$, $g_1^{\nu n}$ probe the
$u$ quark.
In terms of the neutrino structure functions, the unpolarized ratio
$R^\nu = F_1^{\nu p}/F_1^{\nu n}$ is therefore given by:
\begin{eqnarray}
\label{du_nu}
R^\nu &=& { d \over u }\ ,
\end{eqnarray}
while the polarization asymmetries
$A_1^{\nu N} = g_1^{\nu N}/F_1^{\nu N}$ become:
\begin{eqnarray}
A_1^{\nu p} &=& { \Delta d \over d }\ ,
\label{A1nup_qpm}			\\
A_1^{\nu n} &=& { \Delta u \over u }\ .
\label{A1nun_qpm}
\end{eqnarray}
%


\newpage

\begin{table}[h]	
\begin{tabular}{||c||c|c|c|c|c||c||}		\hline\hline
SU(6) rep.	& $^2${\bf 8}[{\bf 56}$^+$]\ \
		& $^4${\bf 10}[{\bf 56}$^+$]\ \
		& $^2${\bf 8}[{\bf 70}$^-$]\ \
		& $^4${\bf 8}[{\bf 70}$^-$]\ \
		& $^2${\bf 10}[{\bf 70}$^-$]\ \
		& total\ \				\\ \hline
$F_1^p$	& $9 \rho^2$
	& $8 \lambda^2$
	& $9 \rho^2$
	& $0$
	& $\lambda^2$
	& $18 \rho^2 + 9 \lambda^2$			\\
$F_1^n$ & $(3 \rho + \lambda)^2/4$
	& $8 \lambda^2$
	& $(3 \rho - \lambda)^2/4$
	& $4 \lambda^2$
	& $\lambda^2$
	& $(9 \rho^2 + 27 \lambda^2)/2$			\\ \hline
$g_1^p$ & $9 \rho^2$
	& $-4 \lambda^2$
	& $9 \rho^2$
	& $0$
	& $\lambda^2$
	& $18 \rho^2 - 3 \lambda^2$			\\
$g_1^n$ & $(3 \rho + \lambda)^2/4$
	& $-4 \lambda^2$
	& $(3 \rho - \lambda)^2/4$
	& $-2 \lambda^2$
	& $\lambda^2$
	& $(9 \rho^2 - 9 \lambda^2)/2$			\\ \hline\hline
\end{tabular}
\vspace*{0.5cm}
\caption{Relative strengths of electromagnetic $N \to N^*$ transitions
	in the SU(6) quark model.  The coefficients $\lambda$ and $\rho$
	denote the relative strengths of the symmetric and antisymmetric
	contributions of the SU(6) ground state wave function.
	The SU(6) limit corresponds to $\lambda = \rho$.\\}
\end{table}

\vspace*{2cm}

\begin{table}[h]	
\begin{tabular}{||c||c|c|c|c|c||c||}			\hline\hline
SU(6) rep.	& $^2${\bf 8}[{\bf 56}$^+$]\ \
		& $^4${\bf 10}[{\bf 56}$^+$]\ \
		& $^2${\bf 8}[{\bf 70}$^-$]\ \
		& $^4${\bf 8}[{\bf 70}$^-$]\ \
		& $^2${\bf 10}[{\bf 70}$^-$]\ \
		& total\ \				\\ \hline
$F_1^{\nu p}$	& $0$
		& $24 \lambda^2$
		& $0$
		& $0$
		& $3 \lambda^2$
		& $27 \lambda^2$			\\
$F_1^{\nu n}$	& $(9 \rho + \lambda)^2/4$
		& $8 \lambda^2$
		& $(9 \rho - \lambda)^2/4$
		& $4 \lambda^2$
		& $\lambda^2$
		& $(81 \rho^2 + 27 \lambda^2)/2$	\\ \hline
$g_1^{\nu p}$	& $0$
		& $-12 \lambda^2$
		& $0$
		& $0$
		& $3 \lambda^2$
		& $-9 \lambda^2$			\\
$g_1^{\nu n}$	& $(9 \rho + \lambda)^2/4$
		& $-4 \lambda^2$
		& $(9 \rho - \lambda)^2/4$
		& $-2 \lambda^2$
		& $\lambda^2$
		& $(81 \rho^2 - 9 \lambda^2)/2$		\\ \hline\hline
\end{tabular}
\vspace*{0.5cm}
\caption{As in Table~I, but for neutrino-induced
	$N \to N^*$ transitions.\\}
\end{table}

\vspace*{2cm}

\begin{table}[h]	
\begin{center}
\begin{tabular}{||c||c|c|c|c|c|c||}		\hline\hline
Model		&    SU(6) \ \
		& no $^4${\bf 10}\ \ \
		& no $^2${\bf 10}, $^4${\bf 10}\ \ \
		& no $S_{3/2}$\ \ \
		& no $\sigma_{3/2}$\ \ \  
		& no $\psi_\lambda$\ \ \	\\ \hline
$R^{np}$	&  2/3          
		& 10/19
		&  1/2
		&  6/19      
		&  3/7        
		&  1/4				\\
$A_1^p$		&  5/9
		&    1      
		&    1       
		&    1
		&    1         
		&    1       			\\
$A_1^n$		&    0         
		&  2/5
		&  1/3
		&    1      
		&    1       
		&    1				\\ \hline
$R^\nu$		&  1/2
		&  3/46
		&    0
		&  1/14
		&  1/5
		&    0				\\
$A_1^{\nu p}$	&--1/3
		&    1
		&   --
		&    1
		&   --
		&--1/3				\\
$A_1^{\nu n}$	&  2/3
		& 20/23
		& 13/15
		&    1
		&    1
		&    1				\\ \hline\hline
\end{tabular}
\vspace*{0.5cm}
\caption{Structure function ratios from quark-hadron duality in SU(6),
	and in various SU(6) breaking scenarios, as described in the
	text.  Note that the ``no $^4${\bf 10}'' and
	``no $^{2,4}${\bf 10}'' scenarios are not consistent with
	quark-hadron duality.\\}
\end{center}
\end{table}

\newpage

\begin{table}[h]	
\begin{tabular}{||c||c|c|c|c|c||c||}		\hline\hline
SU(6) rep.	& $^2${\bf 8}[{\bf 56}$^+$]\ \
		& $^4${\bf 10}[{\bf 56}$^+$]\ \
		& $^2${\bf 8}[{\bf 70}$^-$]\ \
		& $^4${\bf 8}[{\bf 70}$^-$]\ \
		& $^2${\bf 10}[{\bf 70}$^-$]\ \
		& total\ \			\\ \hline
$F_1^p=g_1^p$	& 9 & 2 & 9 & 0 & 1 & 21	\\
$F_1^n=g_1^n$	& 4 & 2 & 1 & 1 & 1 & 9		\\ \hline\hline
\end{tabular}
\vspace*{0.5cm}
\caption{Relative strengths of electromagnetic $N \to N^*$ transitions
	corresponding to $\sigma_{1/2}$ dominance.  These values can be
	obtained from Table~I by adding the $F_1$ and $g_1$
	contributions.\\}
\end{table}

\vspace*{2cm}

\begin{table}[h]	
\begin{tabular}{||c||c|c|c|c|c||c||}		\hline\hline
SU(6) rep.	& $^2${\bf 8}[{\bf 56}$^+$]\ \
		& $^4${\bf 10}[{\bf 56}$^+$]\ \
		& $^2${\bf 8}[{\bf 70}$^-$]\ \
		& $^4${\bf 8}[{\bf 70}$^-$]\ \
		& $^2${\bf 10}[{\bf 70}$^-$]\ \
		& total\ \			\\ \hline
$F_1^{\nu p}
=g_1^{\nu p}$	&  0 &  6 &  0 &  0 &  3 & 9	\\
$F_1^{\nu n}
=g_1^{\nu n}$	& 25 &  2 & 16 &  1 &  1 & 45	\\ \hline\hline
\end{tabular}
\vspace*{0.5cm}
\caption{Relative strengths of $N \to N^*$ transitions in neutrino
	scattering corresponding to $\sigma_{1/2}$ dominance.\\}
\end{table}

\newpage

\begin{figure}[t]       
\begin{center}
\includegraphics{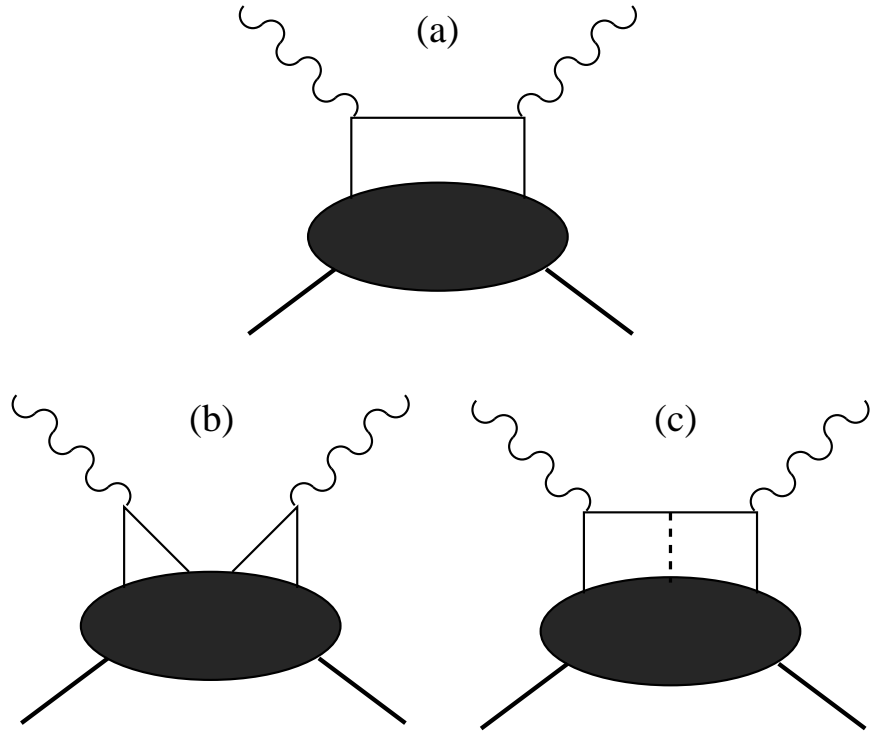}
\vspace*{16cm}
\caption{(a) Leading twist structure function, with photons coupling
	to the same quark; (b) higher twist contributions involving
	coupling to different quarks in the nucleon.}
\end{center}
\end{figure}

\begin{figure}[ht]	
\vspace*{1cm}
\begin{center}
\epsfysize=12cm
\leavevmode
\epsfbox{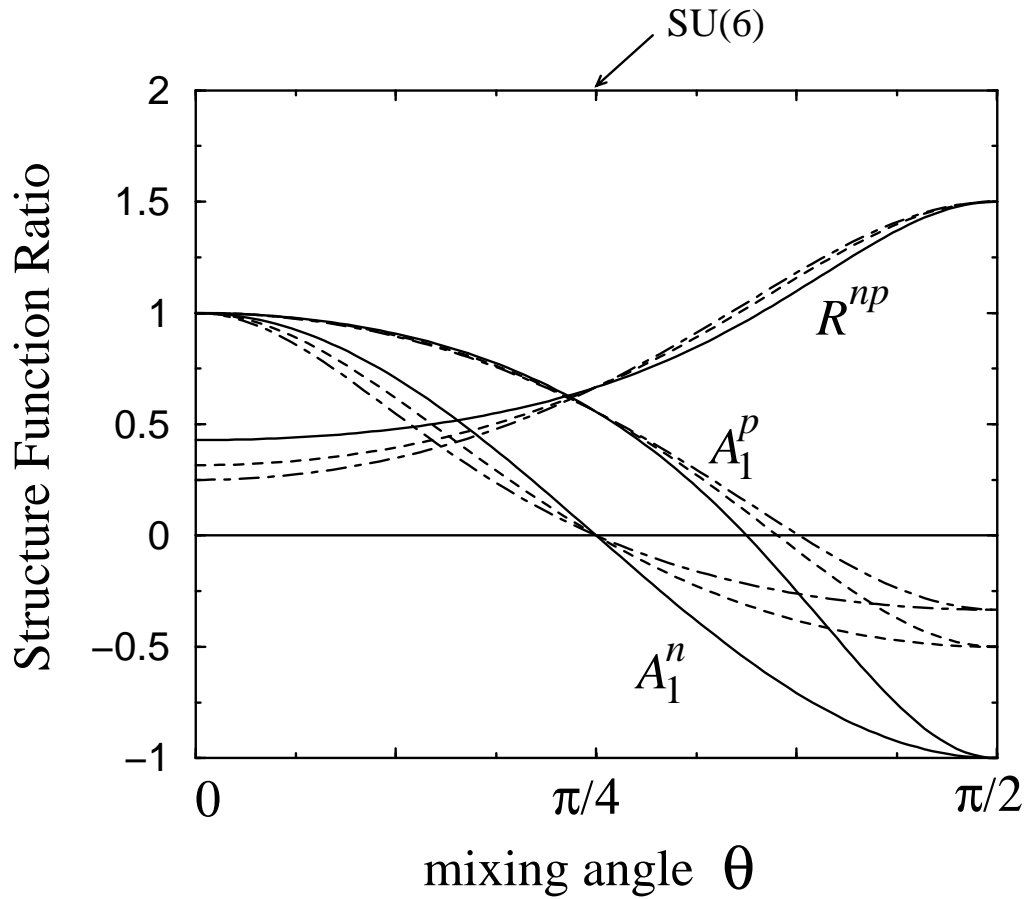}
\vspace*{0.5cm}
\caption{Electromagnetic structure function ratios for different
	combinations of $\sigma_{1/2}$ and $\sigma_{3/2}$ cross sections
	($\theta=\theta_s$, solid), quark spins $S_{1/2}$ and $S_{3/2}$
	($\theta=\theta_h$, dashed), and the symmetric ``$\lambda$'' and
	antisymmetric ``$\rho$'' components of the ground state wave
	function ($\theta=\theta_w$, dot-dashed).  The SU(6) corresponds
	to $\theta=\pi/4$.}
\label{fig:phi}
\end{center}
\end{figure}

\begin{figure}[ht]	
\vspace*{1cm}
\begin{center}
\epsfysize=12cm
\leavevmode
\epsfbox{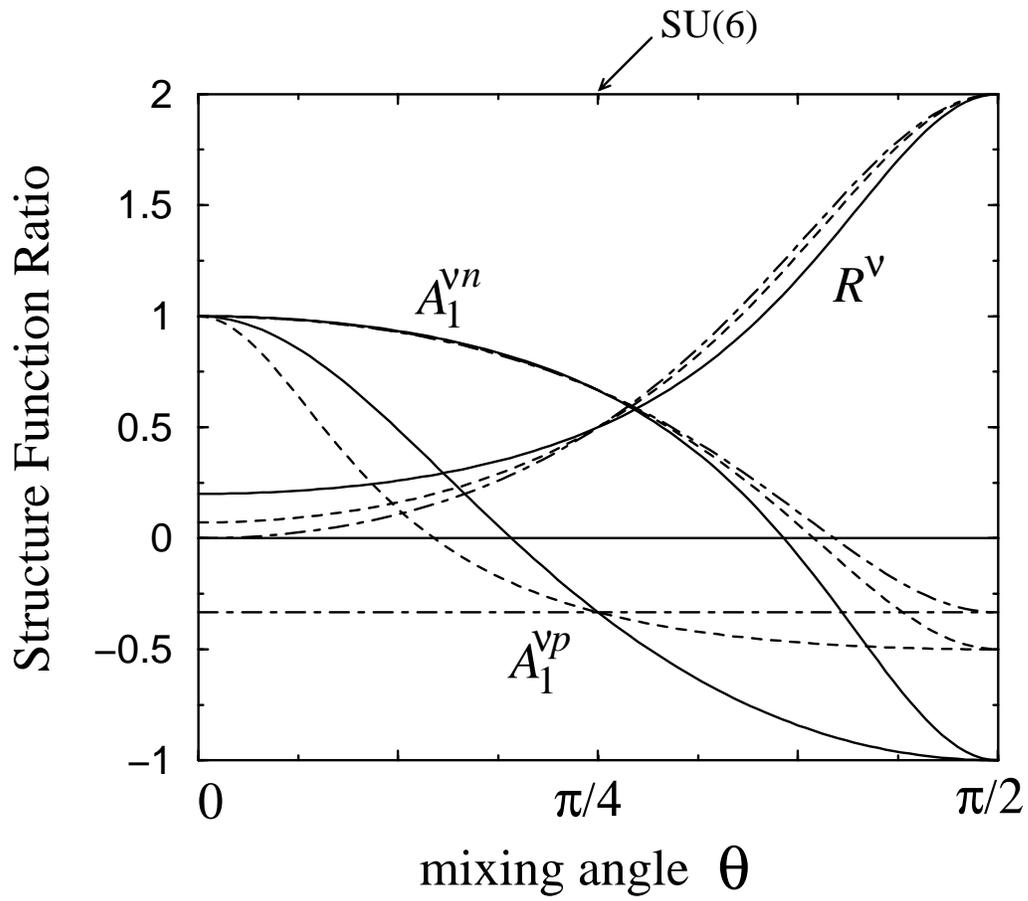}
\vspace*{0.5cm}
\caption{As in Fig.~2, but for neutrino scattering ratios.}
\label{fig:phinu}
\end{center}
\end{figure}

\begin{figure}[ht]	
\vspace*{1cm}
\begin{center}
\epsfysize=12cm
\leavevmode
\epsfbox{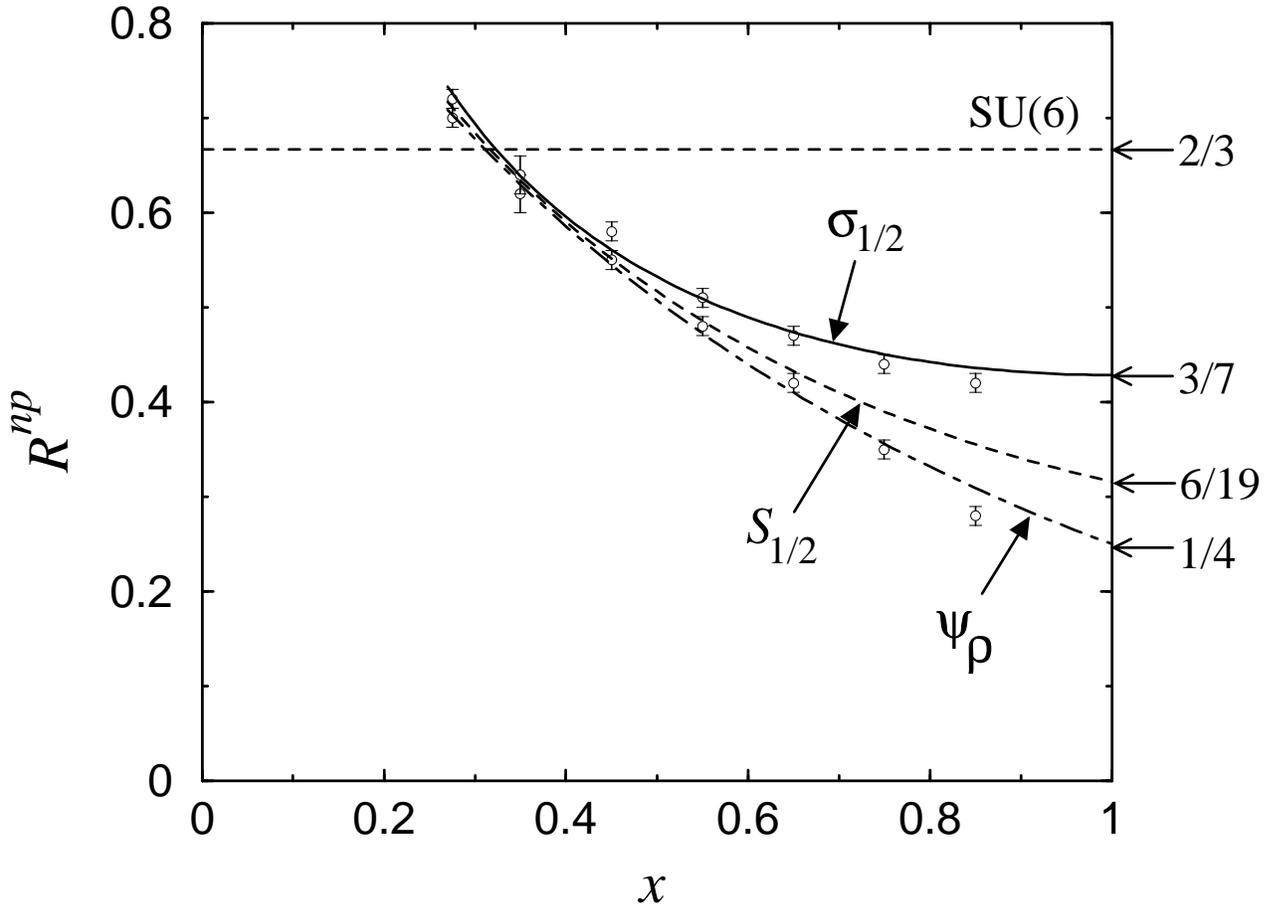}
\vspace*{0.5cm}
\caption{Ratio $R^{np}$ of unpolarized neutron to proton structure
	functions from duality, according to different scenarios of
	SU(6) breaking: helicity $\sigma_{1/2}$ dominance (solid);
	spin $S_{1/2}$ dominance (dashed); $\psi_\rho$ dominance
	(dot-dashed).
	Various theoretical predictions for the $x\to 1$ limit are
	indicated on the ordinate.
	The data are from SLAC \protect\cite{WHITLOW,GOMEZ}, analyzed
	under different assumptions (see text) about the size of the
	nuclear EMC effects in the deuteron \protect\cite{NP}.}
\label{fig:Rnp}
\end{center}
\end{figure}

\begin{figure}[ht]	
\vspace*{1cm}
\begin{center}
\epsfysize=12cm
\leavevmode
\epsfbox{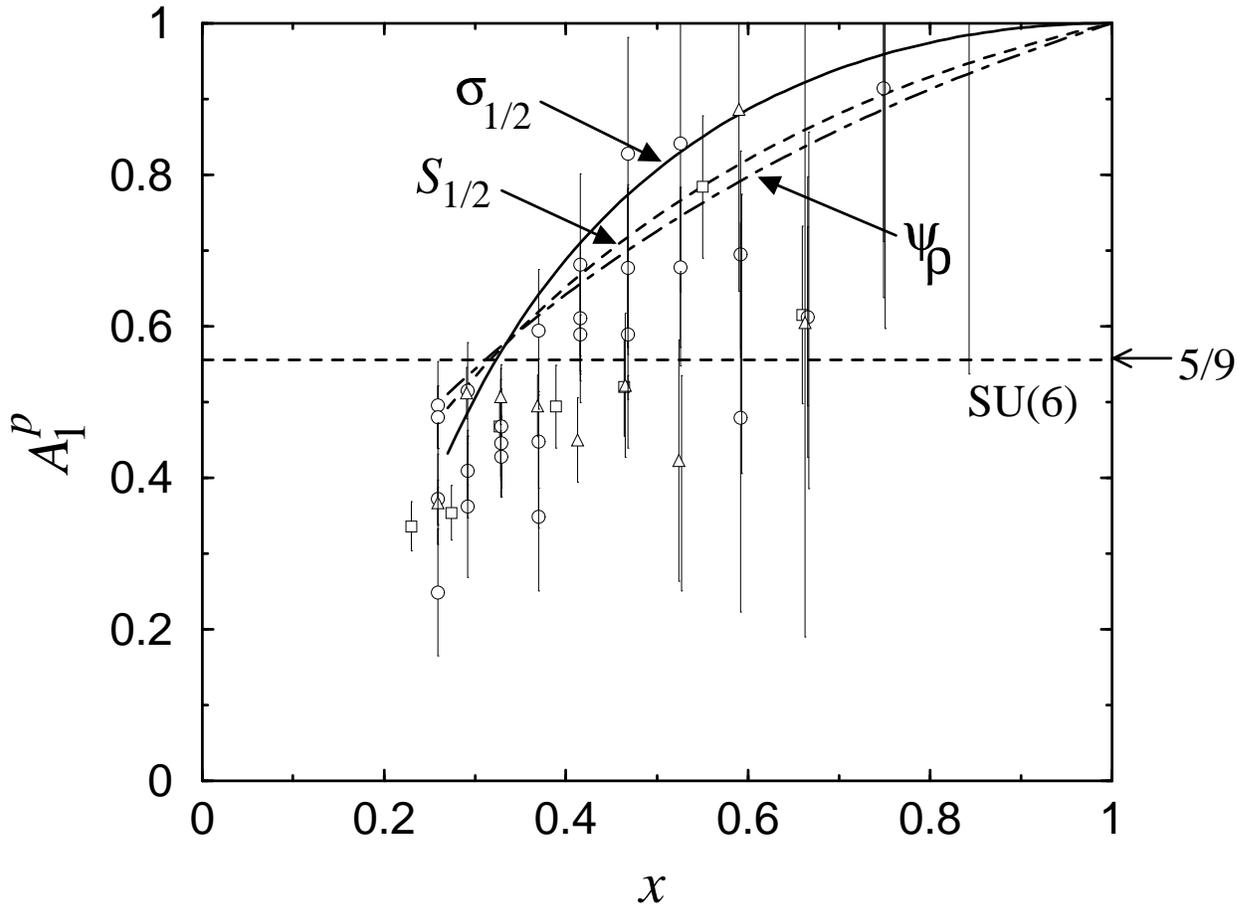}
\vspace*{0.5cm}
\caption{As in Fig.~4, but for the proton polarization asymmetry $A_1^p$.
	The data are a compilation (for $x \agt 0.2$) from experiments at
	SLAC \protect\cite{SLAC}, from the SMC \protect\cite{SMC} and
	HERMES collaborations \protect\cite{HERMES}.}
\label{fig:A1p}
\end{center}
\end{figure}

\begin{figure}[ht]	
\vspace*{1cm}
\begin{center}
\epsfysize=12cm
\leavevmode
\epsfbox{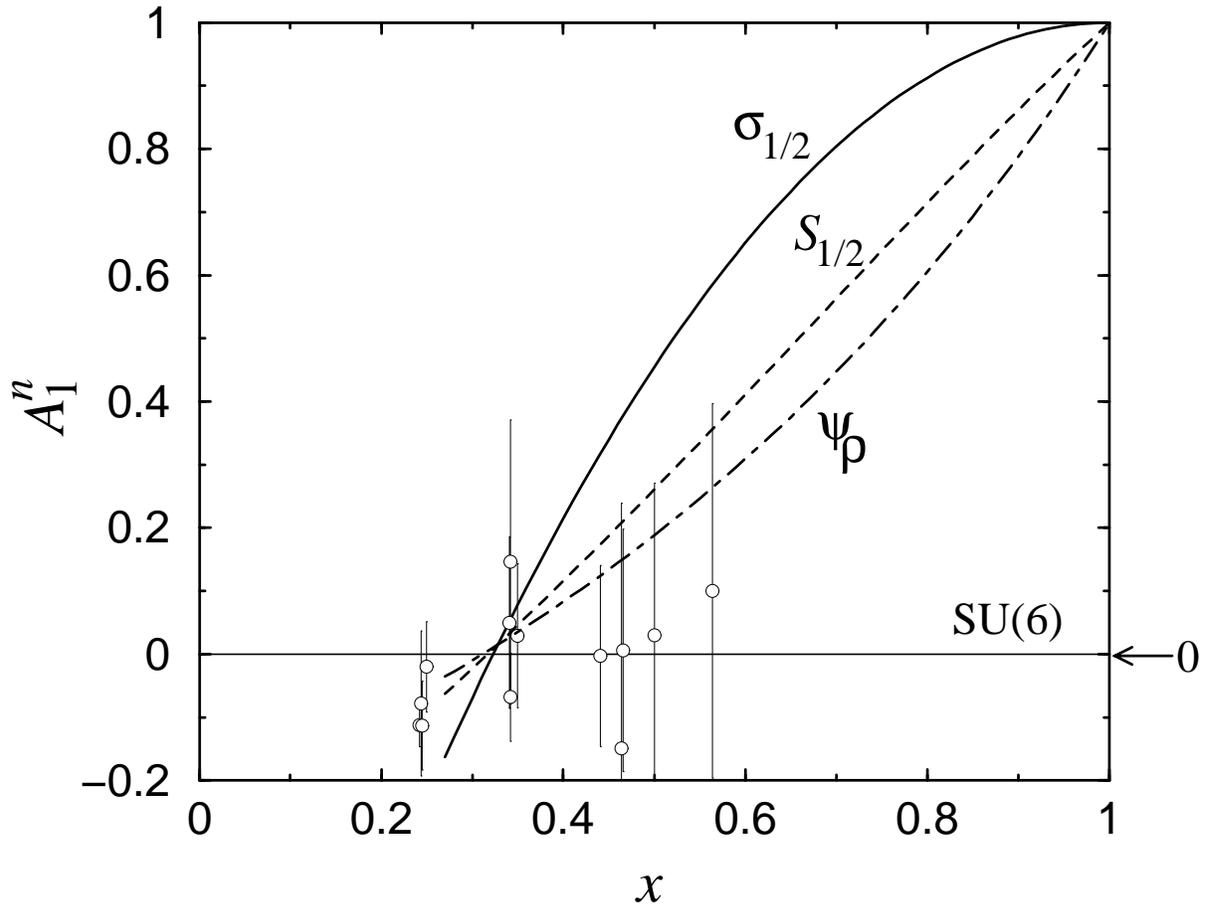}
\vspace*{0.5cm}
\caption{As in Fig.~5, but for the neutron polarization asymmetry
	$A_1^n$.}
\label{fig:A1n}
\end{center}
\end{figure}

\begin{figure}[ht]	
\vspace*{1cm}
\begin{center}
\epsfysize=12cm
\leavevmode
\epsfbox{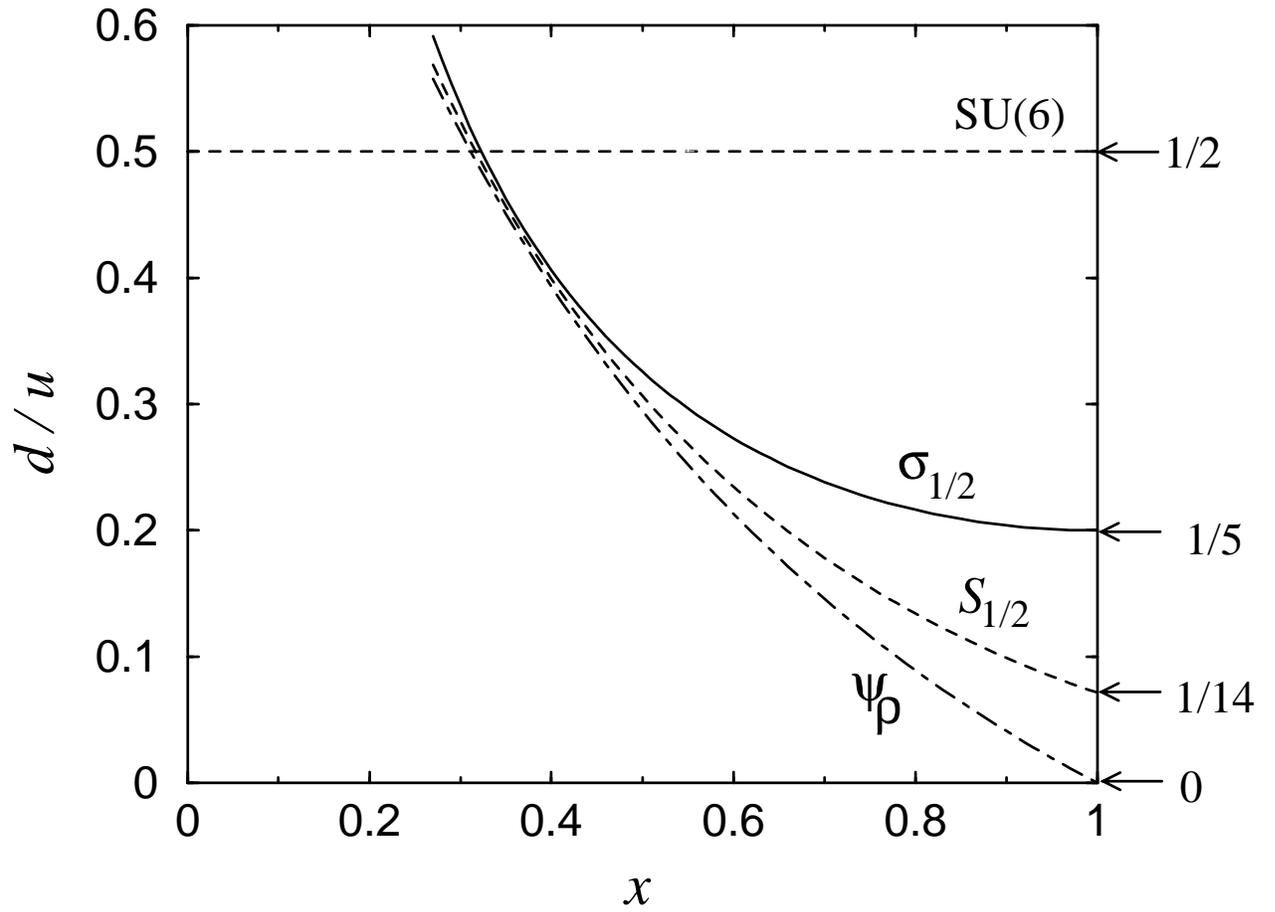}
\vspace*{0.5cm}
\caption{Unpolarized $d/u$ $(= R^\nu)$ ratio in various SU(6) breaking
	scenarios, as described in the text.}
\end{center}
\label{fig:du}
\end{figure}

\begin{figure}[ht]	
\vspace*{1cm}
\begin{center}
\epsfysize=12cm
\leavevmode
\epsfbox{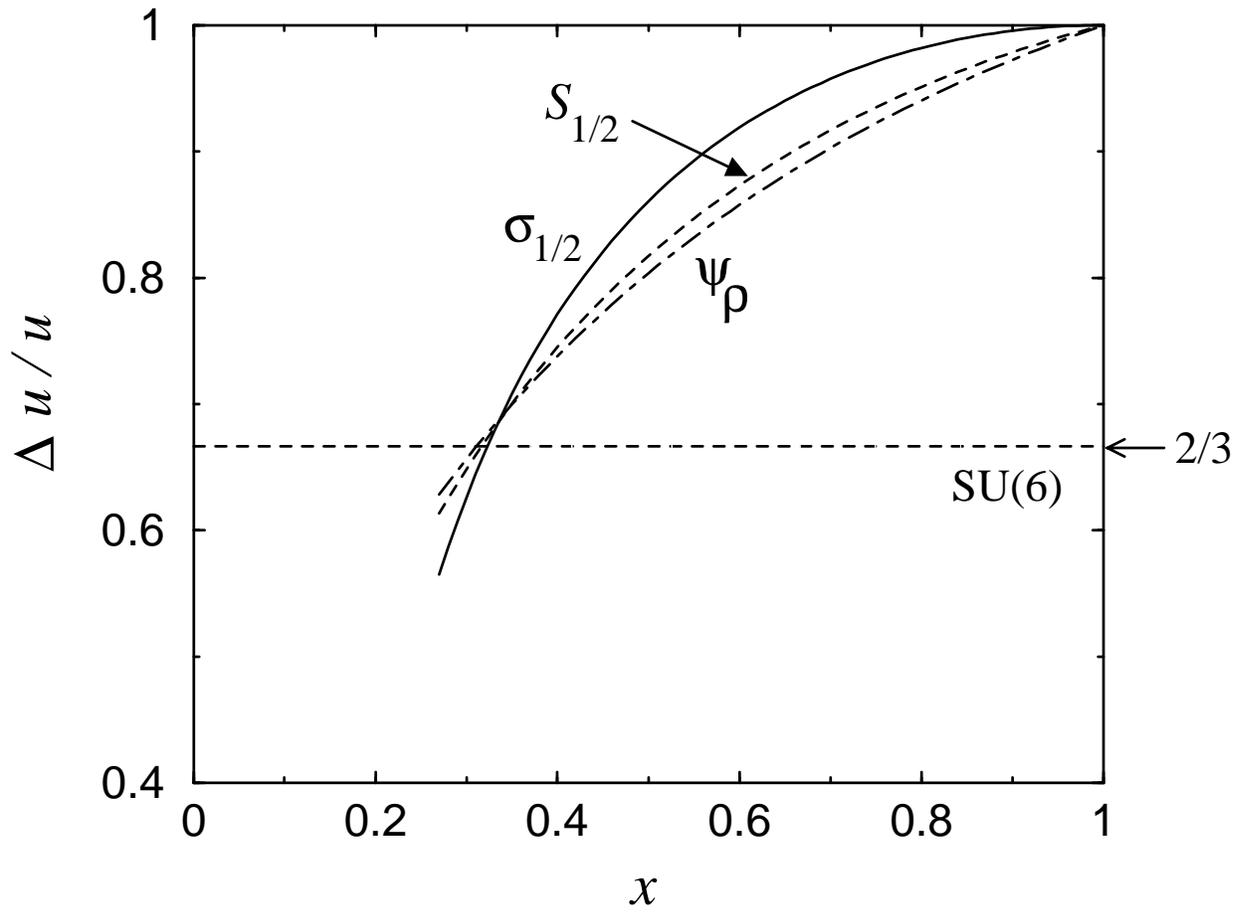}
\vspace*{0.5cm}
\caption{Ratio of polarized to unpolarized $u$ quark distributions,
	$\Delta u/u$ $(= A_1^{\nu n})$, in various SU(6) breaking
	scenarios.}
\end{center}
\label{fig:Duu}
\end{figure}

\begin{figure}[ht]	
\vspace*{1cm}
\begin{center}
\epsfysize=12cm
\leavevmode
\epsfbox{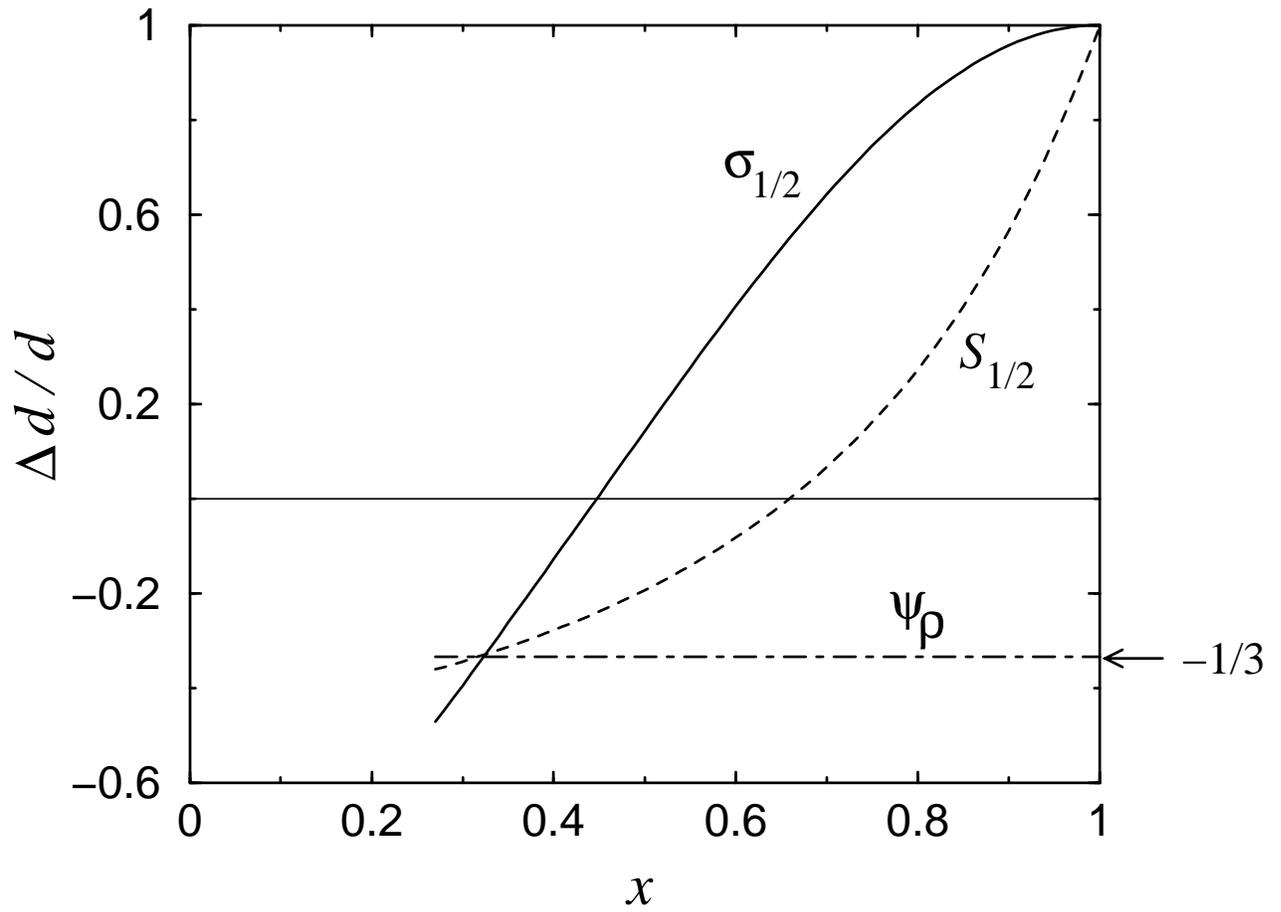}
\vspace*{0.5cm}
\caption{As in Fig.~8, but for the $\Delta d/d$ $(= A_1^{\nu p})$ ratio.}
\end{center}
\label{fig:Ddd}
\end{figure}

\end{document}